\documentclass[12pt,preprint,authoryear]{elsarticle}
\usepackage{natbib}
\usepackage{amssymb}
\usepackage{cases}
\begin{document}

\begin{frontmatter}

\title{Crater Density Predictions for New Horizons flyby target 2014 MU69}

\author{Sarah Greenstreet}

\address{B612 Asteroid Institute, 20 Sunnyside Ave, Suite 427, Mill Valley, CA 94941}
\address{DIRAC Center, Department of Astronomy, University of Washington, 3910 15th Ave NE, Seattle, WA 98195}

\author{Brett Gladman}

\address{Department of Physics \& Astronomy, 6224 Agricultural Road, University of British Columbia, Vancouver, British Columbia}

\author{William B. McKinnon}

\address{Department of Earth and Planetary Sciences and McDonnell Center for Space Sciences, One Brookings Drive, Washington University, St. Louis, MO 63130, United States}

\author{J.~J. Kavelaars}

\address{National Research Council of Canada, Victoria, BC, Canada}

\author{Kelsi N. Singer}

\address{Southwest Research Institute, 1050 Walnut Street, Suite 300, Boulder, CO 80302, United States}

\begin{abstract}

In preparation for the Jan 1/2019 encounter between the New Horizons spacecraft and the Kuiper Belt object 2014 MU69, we provide estimates of the expected impact crater surface density on the Kuiper Belt object.
Using the observed crater fields on Charon and Pluto down to the resolution limit of the 2015 New Horizons flyby of those bodies and estimates of the orbital distribution of the crater-forming projectiles, we calculate the number of craters per unit area formed as a function of the time a surface on 2014 MU69 has been exposed to bombardment.
We find that if the shallow crater-size distribution from 1-15 km exhibited on Pluto and Charon is indeed due to the sizes of Kuiper Belt projectiles, 
2014 MU69 should exhibit a surface that is only lightly cratered below 1 km scale, despite being bombarded for $\sim4$~billion years.
Its surface should therefore be more clearly indicative of its accretionary environment.
In addition, this object may the first observed for which the majority of the bombardment is from exogenic projectiles 
moving less than or near the speed of sound in the target materials, implying morphologies more akin to 
secondary craters elsewhere in the Solar System.
Lastly, if the shallow Kuiper Belt size distribution implied from the Pluto and Charon imaging is confirmed at MU69, then
we conclude that this size distribution is a preserved relic of its state $\simeq$4.5~Gyr ago and provides a 
direct constraint on the planetesimal formation process itself.

\end{abstract}

\end{frontmatter}

\section{Introduction}
\label{Section:intro}

The upcoming flyby of the New Horizons spacecraft by the cold classical transneptunian object (TNO) 2014 MU69 (hereafter referred to as MU69) will offer 
the first up-close look (sub-km resolution) of a small outer Solar System body in its formation environment, and
will provide the first opportunity to observe a high-resolution cratered surface of a TNO other than those in the
Pluto system, imaged by New Horizons in July 2015 \citep{Sternetal2015,Mooreetal2018}.

The classical Kuiper belt appears to be divided into at least two separate inclination components \citep{Brown2001}.  The `cold' classical Kuiper belt \citep{Kavelaars2008} contains the vast majority of the TNOs on low-eccentricity orbits {\it of low inclination} in the semimajor axis range $42.4 < a < 47$~au \citep{petitetal2011}. 
The existence of large separation TNO binaries combined with the low orbital inclination distribution suggests that the cold classicals represent a reservoir of the primordial proto-planetary disk beyond 40~au \citep{Parker2010}. 
This the only sub-population thought to have been in place since the formation of the Solar System.
The low orbital inclination and eccentricity of MU69's orbit \citep{Porteretal2018} place this object firmly within this region and thus
MU69 is likely to have formed at this large heliocentric distance and be the most primitive outer Solar System body yet visited.
There is considerable structure in the $a$ and eccentricity ($e$) space of the cold population, and it is has been split \citep{petitetal2011} into a `stirred' component crossing the whole cold $a$ range, and a much more confined `kernel' of cold objects near $a=44$~au and $e\simeq0.05$; MU69 is consistent with being in this kernel.

The other Kuiper Belt sub-populations (characterized by broader orbital inclination distributions) have been postulated
to have formed closer to the Sun and transplanted to their current location in the final
stages of planet formation \citep[e.g.][]{Levisonetal2008}.
The postulated 500~Myr delay in the instability in the often-cited Nice Model \citep{Gomesetal2005} has been retreated from
\citep{Mann2018}; see \citet{Nesvorny2018} for a recent review.
The desirable implantation properties for creating Kuiper Belt structure, however, still hold even if the instability process 
occurs early in Solar System history \citep{Gladmanetal2012}.
In the scenario that MU69 has always been near its current orbit, a hypothetical metastable phase (which was of interest for
Pluto's early cratering since Pluto is part of the metastable population)  is irrelevant because during the early
phase the population closer to the Sun is not cratering MU69.
In either case, the `hot' populations are ultimately emplaced during a relatively brief ($<1$\% of the Solar System's age)
period in which the cold population must not be dynamically excited \citep[eg.,][]{Dawson2012}.
While we thus concentrate our calculations on the last 4 Gyr of bombardment, we argue below that the
{\em brief} phase in which the objects where scattered out through the Kuiper belt, with some small fraction becoming the hot classical Kuiper belt we see today, cannot contribute a large change in MU69 cratering,
especially given that we find that the cold populations dominate the impactor flux.

The observed crater density on Pluto and Charon by the New Horizons spacecraft during its flyby of the Pluto system in July 2015 showed a change to a shallower size distribution \citep{Singeretal2019}.
We assume here that this transition is caused by a similar paucity of small impactors, and combine the derived
impactor size distribution
with a Kuiper belt orbit distribution model
to compute the impactor flux, impactor speed
distributions, and resulting crater formation rates on MU69.

\section{The impactor size distribution}
\label{Section:uncert_sd}

\begin{figure}[h!]
\includegraphics[scale=0.50]{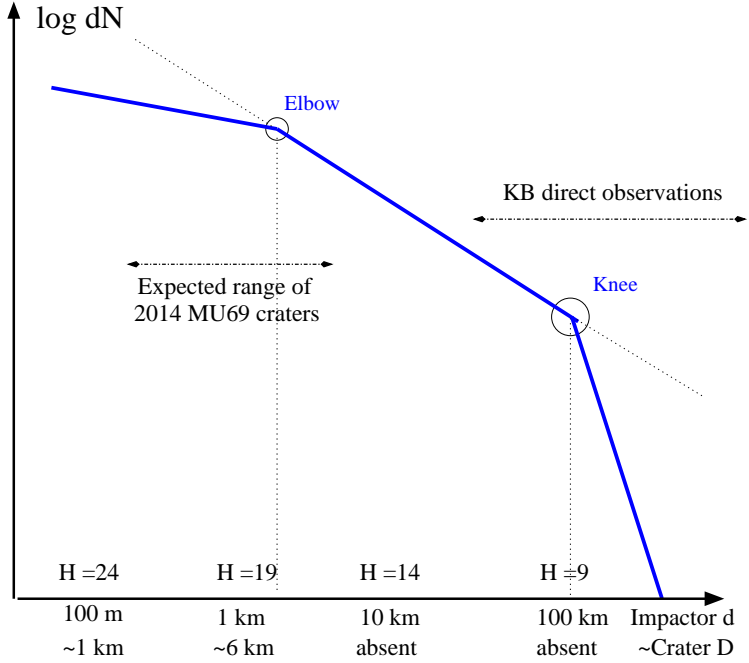}
\caption{Cartoon schematic of the $H_{g}$-magnitude differential size distribution. The Kuiper belt observations are well calibrated down to $H_{g}\approx10$, but smaller than the knee at $H_g\sim9$ (open circle) the size distribution flattens from a slope of $\alpha\simeq0.8$ (not relevant to this paper) down to a shallower $\alpha\simeq0.4$. 
\citet{Singeretal2019} find the impactor size distribution from craters observed on Charon requires a second change
to a slope of  $\alpha\simeq0.15$ for $d\lesssim2$~km.
For reference, impactor diameters $d$ are converted to rough MU69 crater diameters $D$ assuming a common impact speed of 300~m/s. 
The expected range of craters observable by New Horizons covers $D\approx200$~m to $D\approx20$~km 
(created by cold-belt impactors ranging from $d\approx0.02-3$~km).
Notice that the ratio of crater to impactor diameters is smaller than usual due to the low impact speeds.
}
\label{Fig:size_dist}
\end{figure}

Calculating crater formation rates requires a knowledge of the impactor population's orbital distribution
(which determines the impact probabilities and impact speed distribution) and its diameter $d$ distribution
in order to compute the scale of the resulting craters of diameter $D$.
The differential number of objects $dN$ as a function of absolute $H$-magnitude behaves locally as
$dN\propto10^{\alpha H}$, where the exponential index $\alpha$ is also referred to as the logarithmic ``slope" (hereafter referred to simply as the slope).
This corresponds to a cumulative power law distribution in the projectile diameter with $N(<d)\propto d^{-5\alpha}$.
The conversion for objects with absolute g-band magnitude $H_{g}$ and scaling to an albedo $p$ of 5\% is
\begin{equation}
d\backsimeq100~{\mathrm {km}}\;\sqrt{{\frac{0.05}{p}}}\;10^{0.2\left(9.16-H_{g}\right)}
\label{eq:DtoH}
\end{equation}
where we see that $H_g$=9.16 corresponds to 100 km.
We have chosen to make a single pre-encounter crater-density prediction,
based on telescopic observations and the results of the New Horizons flyby of Pluto and Charon. 
We use one size distribution 
(Fig.~\ref{Fig:size_dist}) 
for all of the various Kuiper Belt sub-populations.
The steep observed Kuiper Belt size distribution for TNOs $d>100$~km rolls at this diameter to a 
``knee" size distribution, as used in \citet{Greenstreetetal2015,Greenstreetetal2016}; the slope
$\alpha$=0.4 below this knee extends down to another slope change at a scale which 
\citet{Singeretal2019}
denote as the ``elbow'', visible on both Pluto and Charon.
On Charon, this corresponds to a crater size of $D\approx13$~km, which for typical Charon impact speeds
implies projectile $d\approx2$~km and thus $H_g\approx17.5$.
We connect the $H_g<9.16$ population estimates from \citet{petitetal2011} and \citet{Gladmanetal2012} to 
the $H_g<17.5$ scale by using a multiplicative factor of $10^{\alpha \Delta H}=10^{((0.4*8.5)-(0.8*0.16))}\approx1,900$.
Below this scale Pluto and Charon show a very shallow distribution with $\alpha \simeq 0.15$, and
we assume this slope is that coming from the impactor size distribution and directly map it
back to the projectiles.
Fig.~\ref{Fig:size_dist}'s horizontal axis also shows the expected crater diameter caused by a typical 
cold-population impactor striking MU69 (see below).
If this crater size distribution is in fact present on MU69 it will be a dramatic confirmation
of a shallow Kuiper Belt size distribution in the roughly 0.1--2~km diameter range.

\section{Methods}
\label{Section:methods}

The methods for computing the current impact fluxes and cratering rates onto MU69 are identical to those in \citet{Greenstreetetal2015,Greenstreetetal2016}, to which we refer the reader for the details; below we only 
mention any deviations from the methods used in the previous paper.

\subsection{Kuiper belt population models}
\label{Section:pop_models}

MU69 sits in the heart of the cold classical Kuiper Belt's kernel 
with the encounter target having $a=44.2$~AU, $e=0.04$, and heliocentric J2000 orbital inclination $i=2.4^o$ \citep{Porteretal2018}. 
We adopted the same orbital distributions and $N(H_g<9.0)$ population estimates for the various Kuiper Belt sub-populations as 
before
and computed the impact fluxes onto MU69. 
We added to the analysis the population of TNOs in the 7:4 mean-motion resonance with Neptune
(whose population is small and was thus not used for the Pluto/Charon analysis),
because the resonant semimajor axis of $a\simeq43.6$~AU is close to MU69 and the impact probability
per particle is enhanced,
thus contributing a non-negligible fraction of the total impact flux onto MU69.

As in \citet{Greenstreetetal2015,Greenstreetetal2016}, we account for the slow decay of each projectile
sub-population over the last 4 Gyr when we compute the total impact flux.
For the 7:4 resonance we adopt the same decay as for the 2:1 objects found in \citet{TiscarenoMalhotra2009}.
These enhancements back in time are modest except for the scattering TNO population, but we show below that this
population is only a tiny contributor to MU69 cratering.


\subsection{\"{O}pik collision probability code}
\label{Section:pcol}

The \"{O}pik collision probability code used in this study is the same as used previously, with the only change being the new target body. 
MU69's gravitational focusing is negligible, but is included.
Unlike computing the impact flux onto Pluto from the various sub-populations, 2014 MU69 is not located within a mean-motion resonance 
nor is it undergoing Kozai oscillations.
It was thus not necessary to provide a correction factor to the collision probabilities provided 
by the \"{O}pik collision probability code to take into account how the dynamics allowed Pluto to avoid orbital
intersections with some portions of some sub-populations.
Our code provides the impact probabilities for the ensemble of objects in each orbit and the probability-weighted
impact speed distribution onto MU69.

\begin{figure}[h!]
\centering
\includegraphics[scale=0.5,angle=-90]{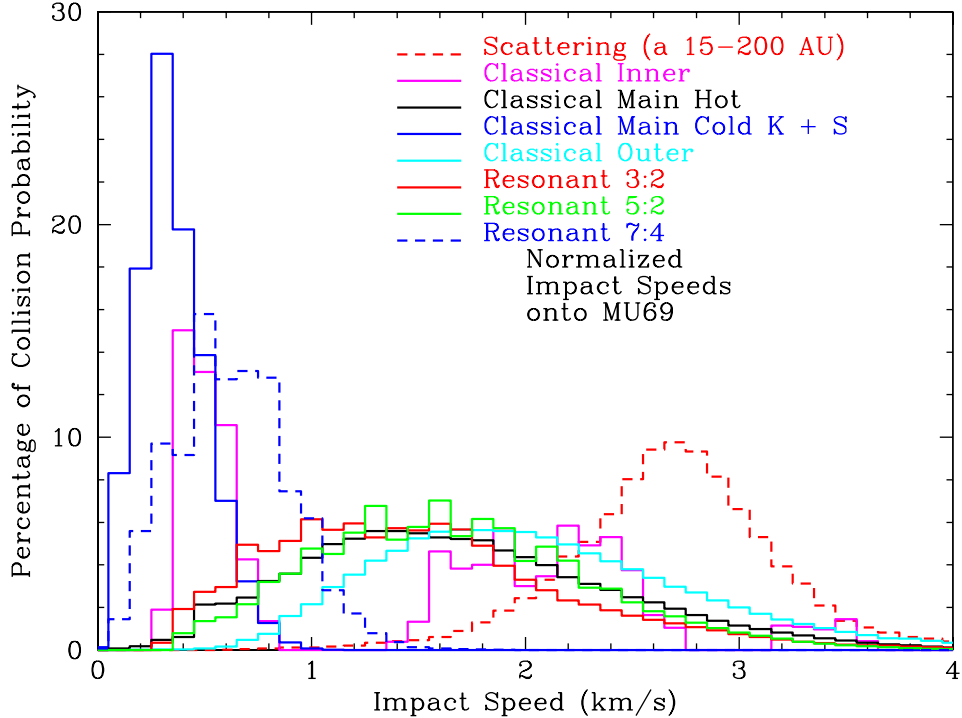}
\caption{Impact velocity spectrum onto MU69 for each Kuiper Belt sub-population.
Each sub-population's distribution is separately normalized.
The cold main classical impact velocity distribution includes both the stirred and kernel sub-components, 
because their speed distributions are similar.
The bimodal nature of the (nearly negligible) inner classical belt is due to a gap in that population's
orbital inclination distribution \citep{petitetal2011}.
Almost all impactors on MU69 are travelling less than the speed of sound in coherent water ice. 
}
\label{Fig:impact_speeds}
\end{figure}

The impact speed distribution is remarkable (Fig.~\ref{Fig:impact_speeds}).
Essentially all impacts onto MU69 are less than the $\sim$4~km/s \textit{p} wave speed in solid water
ice; thus if MU69 is a coherent body these impacts
are subsonic, where the crater scaling laws are less well-established
(see discussion below and in \citet{Singeretal2013}).
The most extreme behavior is that, because of their similar orbits to MU69, the cold classicals have 
tiny impact speeds, peaking around only 300~m/s, similar to the \textit{p} wave speed in unconsolidated sands - which may be a better structural analog to MU69's surface.
While such speeds are above MU69's escape speed, they are essentially unknown for primary
projectiles in the Solar System and would only be seen elsewhere in the context of non-escaping,
secondary projectiles.
Fig.~\ref{Fig:impact_speeds} shows, as expected, the 7:4 resonant objects have a range of encounter speeds including serendipitous close encounters 
with MU69 at similarly low velocities due to their very similar semimajor axes.
Unsurprisingly the scattering objects, which have large semimajor axes, peak at the highest impact velocity ($\approx3$~km/s) 
of the sub-populations. 
The remaining sub-populations peak at similar impact speeds near $\approx1.5$~km/s.

\subsection{Impact rates onto MU69}
\label{Section:impact_rate}

Table~\ref{Table:impact_rate} shows the calculated \"{O}pik collision probabilities (/yr/TNO) onto MU69 for each Kuiper Belt sub-population
(this collision probability is independent of the projectile size).
This is converted to an impact rate by multiplying by the estimated corresponding population at some scale.
To pick a single number, we have chosen to tabulate the rate of projectiles of the size scale of the elbow or larger
($H_g<17.5$ or $d\gtrsim2$~km).
This reveals that the impact rate is dominated by the three components of the main classical belt, which together supply
(10+58+18)=86\% of the projectiles that have struck MU69 over the last 4~Gyr.
The total impact rate of projectiles larger than the elbow is (perhaps surprisingly) low at only 0.017/Gyr; it is thus
likely that over the entire post-accretionary bombardment history no projectile with $d\gtrsim1$~km has struck MU69 (we assume an effective target radius of 10~km).
We thus predict that large craters will be rare or absent; we quantify this below.
The impact flux from impactors smaller than the elbow will of course be larger due to their increasing numbers as diameters drop,
but with a shallow slope they do not become numerous quickly. 
Our calculations of crater production that follow peform the full integration over the size and speed distributions, and are
thus better than quoting a single number.

Another remarkable result shown in Table~\ref{Table:impact_rate} is that the scattering population is a fractionally negligible
addition to the impact rate.
The resonant populations are each small contributors.
The large 3:2 population (which was the most important of the sub-populations for Pluto cratering) has a low impact
probability per particle due to the large mutual orbital inclination with MU69.
The enhancement of the 7:4 population (by about a factor of two per particle) is evident, but the four
most important resonant sources together provide $<10$\% of the impactor flux.
The classical inner population is essentially negligible because many of its members do not intersect
the MU69 orbit.
Finally, the population of non-resonant TNOs with semimajor axes beyond the 2:1 resonance (the `Classical Outer' population)
contributes about 6\% of the impactors; this population has many orbits with perihelion just inside MU69's
perihelion and as Fig.~2 shows these impactors have median impact speeds $U\sim2$~km/s.

\begin{table}[h!]
\begin{center}
\tiny
\tabcolsep=0.11cm
\begin{tabular}{| c | c || c | c | c || c | c |}
\hline
\bf{Kuiper Belt} & \bf{$H_{g}<17.5$} & \bf{\"{O}pik} & \bf{$d\gtrsim2$~km} & \bf{\% of} & \bf{$D>1$~km} & \bf{\% of}\\
\bf{Sub-Population} & \bf{Population} & \bf{Impact} & \bf{Impact} & \bf{Total} & \bf{Cratering} & \bf{Total} \\
\bf{Type} & \bf{Estimate} & \bf{Probability} & \bf{Rate} & \bf{Impact} & \bf{Rate} & \bf{Cratering} \\
\bf{} & \bf{} & \bf{(/yr/TNO)} & \bf{(/Gyr)} & \bf{Rate} & \bf{(/Gyr)} & \bf{Rate} \\
\hline
S.O. ($15$~AU~$\leq$ & 2.0e7 & 3.4e-21 & 6.8e-5 & 0.4 & 7.3e-3 & 0.7 \\
$a\leq200$~AU) &  &  &  &  &  & \\
\hline
S.O. ($a>200$~AU) & 1.8e8 & 1.0e-22 & 1.8e-5 & 0.1 & 2.1e-3 & 0.2 \\
\hline
Classical Inner & 5.5e6 & 5.6e-22 & 3.1e-6 & 0.0002 & 2.6e-4 & 0.0002 \\
\hline
\bf{Classical Main H} & 6.5e7 & 2.4e-20 & 1.6e-3 & \bf{9.5} & 1.4e-1 & \bf{13.2} \\
\hline
\bf{Classical Main S} & 1.4e8 & 7.0e-20 & 9.8e-3 & \bf{58.0} & 5.4e-1 & \bf{50.7} \\
\hline
\bf{Classical Main K} & 3.8e7 & 7.8e-20 & 3.0e-3 & \bf{17.8} & 1.6e-1 & \bf{15.0} \\
\hline
Classical Outer & 1.5e8 & 6.8e-21 & 1.0e-3 & 5.9 & 9.9e-2 & 9.3 \\
\hline
Resonant 3:2 & 2.5e7 & 2.7e-20 & 6.8e-4 & 4.0 & 5.8e-2 & 5.4 \\
\hline
Resonant 2:1 & 7.0e6 & 2.3e-20 & 1.6e-4 & 0.9 & 1.4e-2 & 1.4 \\
\hline
Resonant 5:2 & 2.3e7 & 1.2e-20 & 2.8e-4 & 1.7 & 2.6e-2 & 2.4 \\
\hline
Resonant 7:4 & 5.5e6 & 5.0e-20 & 2.8e-4 & 1.7 & 1.8e-2 & 1.7 \\
\hline
\bf{Total} &  &  & \bf{0.017} & \bf{100.0} & \bf{1.1} & \bf{100.0} \\
\hline
\end{tabular}
\end{center}
\caption{\"{O}pik collision probability calculations, $d\gtrsim2$~km impact rates, and $D>1$~km cratering rates onto MU69 for each Kuiper Belt sub-population. Population estimates are for $H_{g}<17.5$ (diameter $d\gtrsim2$~km, for a g-band albedo $p=5\%$), which is the location of the elbow. Impact probabilities are (/yr/object). Impact rates are (/Gyr) and determined using the number of $H_{g}<17.5$ objects in each sub-population assuming an effective target radius near 10~km (the stellar occultation silhouette is 20 x 35~km, but irregular \citep{Mooreetal2018}). Cratering rates for $D>1$~km are (/Gyr) of bombardment.}
\label{Table:impact_rate}
\end{table}

Our derived impact fluxes, when expressed as impacts per year per km$^2$ of target (assuming an effective spherical radius of $r\approx10$~km), 
are about twice that onto Pluto, reflecting the higher spatial density of the MU69 environment.
This relative flux is in agreement with JeongAhn et al. (2018, to be submitted)
who also conclude that the cold classicals should provide $\sim$80\% of the impacts. 


\subsection{Cratering rates onto MU69}
\label{Section:crater_rate}

For the purposes of this paper we assume that MU69 is an unconsolidated, low-density body. 
We convert between crater diameters $D$ and projectile diameter $d$ via
\begin{equation}
D=8.9\left({\frac{U_{km/s}^{\; 2}}{g_{cm/s^2}}}\right)^{0.170}\left({\frac{\delta}{\rho}}\right)^{0.333} d_{km}^{\ 0.830}~{\mathrm {\; \; \; km}}
\label{eq:scaling_law}
\end{equation}
where $U$ is the impact velocity (in km/s), $d$ is the impactor diameter, and the gravitational acceleration on the surface of MU69 is $g\approx0.3$~cm/s$^{2}$. 
We assume both impactor $\delta$ and target $\rho$ (at surface) densities are $\delta=\rho=1.0$~g/cm$^{3}$, although
the value is unimportant as long as they are the similar.

The coefficients in Eq.~(\ref{eq:scaling_law}) are appropriate to unconsolidated sand or the regolith in the gravity regime \citep{Singeretal2013}, and are based on well-established scaling from laboratory and numerical simulations \citep{Holsapple1993,HousenHolsapple2011, Greenstreetetal2015, Greenstreetetal2016, Singeretal2019}. This procedure should provide a reasonably accurate estimate of crater size (to within +/- 50\%), provided neither the impact speed nor the cratering efficiency (excavated crater mass/impactor mass) are too low.  
Well-defined secondary craters are seen on icy satellites for speeds down to $\approx200$~m/s, so we anticipate 
that most TNO impacts onto MU69 will form craters;
the lowest-speed impactors may be accretionary, however, not crater-forming.
Despite the difference of physical regime, Eq.~\ref{eq:scaling_law} gives quite similar results to those obtained if
we use the scaling law we deployed in \citet{Greenstreetetal2015} appropriate for non-porous coherent surfaces; when
propagated through the entire analysis the resulting crater densities presented below are nearly identical.
Thus, uncertainties in the scaling law is of little consequence compared to the much larger uncertainties
that we expect to arise because of the sparse crater statistics.

\begin{figure}[h!]
\centering
\includegraphics[scale=0.5]{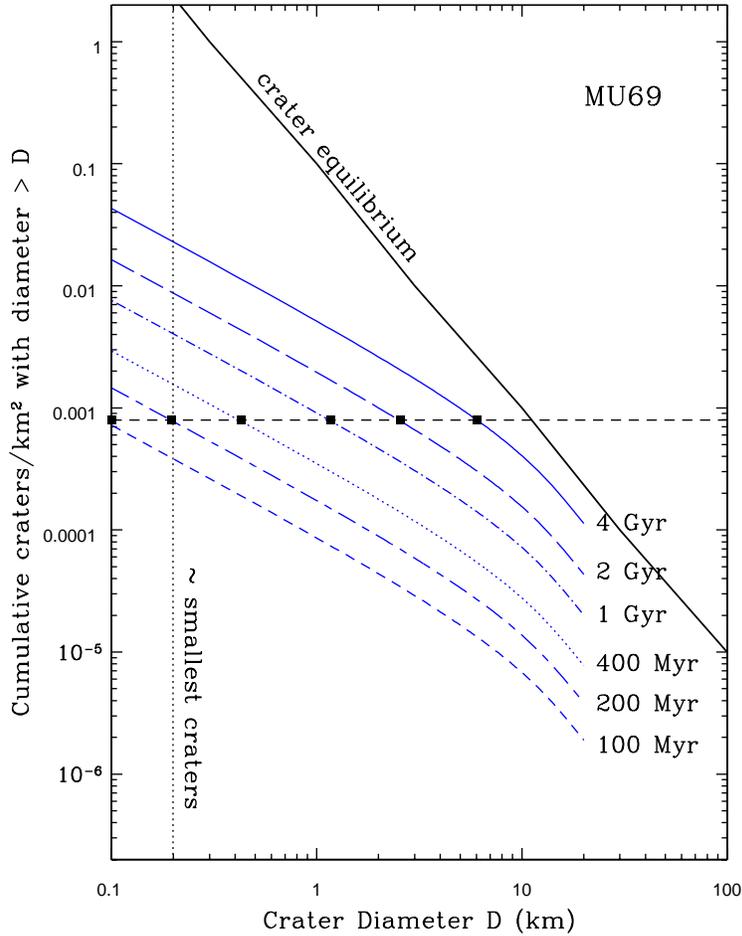}
\caption{Logarithm of crater density (\#~craters/km$^{2}$) larger than a given crater diameter $D$ on MU69's surface versus the logarithm of the crater diameter for an impactor size distribution with a both a knee (not visible in this plot) and elbow (shown as the gradual slope change near $D\approx10$~km) at various surface ages. The solid black line is the crater equilibrium curve from \citet{Melosh1989}. The horizontal line at $\approx10^{-3}$ corresponds to 1 crater/MU69 surface, and the vertical line is an estimate for the smallest craters that will be visible in the best MU69 images. The black squares at the 1 crater/MU69 surface line correspond to the black squares in Figure~\ref{Fig:r_plot}. The change in slope at $D\approx10$~km corresponds to the elbow break in the impactor size distribution, but will not be visible in the crater data; note that this change is gradual because of the large fractional spread in impact speeds $U$ present.}
\label{Fig:num_craters_diameter}
\end{figure}

First note that for the tiny $U\simeq0.3$~km/s, craters are only $D\simeq7$~km for a $d=1$~km projectile, and thus
the crater/projectile ratio is somewhat smaller than the typical 10-20 common elsewhere in the 
solar system.
By integrating the crater production over all speeds of all sub-populations, we derive the crater
production rates shown in Fig.~\ref{Fig:num_craters_diameter} for various bombardment time 
scales.

Although a single number is of limited utility, Table~\ref{Table:impact_rate} also lists the rate of production
of impact craters larger than an arbitrarily-chosen size of $D>1$~km from each projectile
sub-population.
This crater formation rate is low.
Accounting for all the projectiles, the expectation is a modest 5 craters with 
$D>$1~km on MU69 after bombardment for the Solar System's age.
Because of the low impact speeds from the cold classical sources, their dominance is 
somewhat diminished in terms of fraction of the crater production above a given
$D$ limit, as their craters are smaller than those produced by the higher-speed populations; 
the cold population still manages to contribute three-quarters
of the $D>1$~km production.

Fig.~\ref{Fig:num_craters_diameter} provides a much more complete view of our results in
the form of predicted cumulative crater density curves, for several bombardment 
durations (that is, ages since a surface was last reset by geologic processes).
The horizontal line corresponds to 1 crater on the surface.
The conclusion is dramatic: MU69 should be lightly cratered near the resolution limit, 
for despite extremely low gravity on MU69, 
the paucity of small projectiles and their slow impact speeds produce few 
craters with $D\sim200$~m.
Even more spectacularly, given that the surface area of MU69 is only $\sim$1000 -- 2000~km$^2$,
the 4~Gyr bombardment predicts only $\sim$25--50 craters to be
above the effective $\sim200$~m resolution limit of MU69, and note that only half of the body is 
likely to get this highest-resolution data.
As long as craters are visible, however, this predicted level of cratering would provide strong
confirmation that the paucity of Pluto/Charon craters smaller than the elbow scale 
(at crater $D<13$~km for Pluto/Charon or impactor $d\lesssim2$~km) is
indeed a feature of the projectile size distribution.
At the large end, bombardment over the entire age of the Solar System will on average provide
a crater with a maximum diameter of order 6~km (a larger crater could be possible due
to Poisson statistics).

\begin{figure}[h!]
\centering
\includegraphics[scale=0.5]{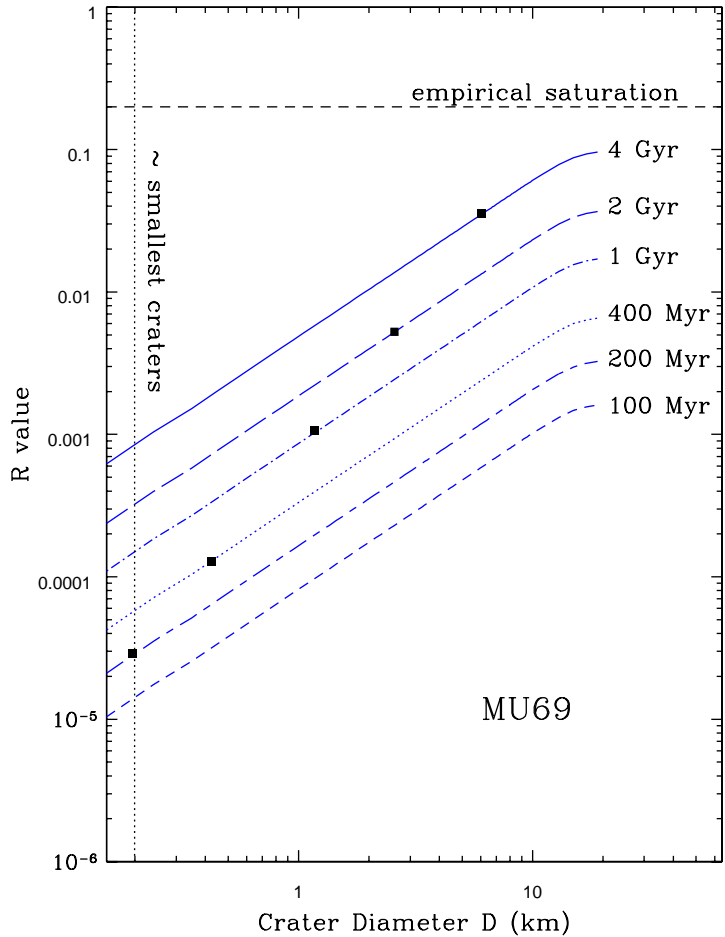}
\caption{Relative crater frequency plot of the same information in Figure~\ref{Fig:num_craters_diameter}. 
The black squares correspond to 1 crater/MU69 surface on the cumulative plot (Figure~\ref{Fig:num_craters_diameter}), so craters larger than the dots will likely not be visible on MU69's post-accretionary terrains, except by statistical fluctuation. 
The fact that R values rise for increasing $D$ is due to the relatively shallow
projectile size distribution implied by the Pluto/Charon surfaces.
Assuming MU69 preserves the craters integrated over all of Solar System history,
this level of cratering in the visible portion will be sparse statistically.
}
\label{Fig:r_plot}
\end{figure}

Fig.~\ref{Fig:r_plot} displays our results in `R-plot' formulation, which can
be thought of as roughly the fraction of the surface covered in craters at
each crater scale $D$. 
The rise with increasing diameter over the visible range would be very 
characteristic of the projectile population derived from THE Pluto/Charon crater
data.
If the (probably sparse) crater counts do indeed match these predicted values,
it would strongly argue that the current Kuiper belt size distribution is
extremely shallow in the 100~m to 1~km diameter range probed by MU69 and
the Pluto-Charon crater counts.
This would then offer the strong possibility that the current Kuiper
Belt retains the primordial planetesimal-building size distribution.

\section{Discussion and Conclusion}
\label{Section:discussion}

Based on the above results, we conclude that bombardment over the entire age of the Solar System
is insufficient to more than modestly crater 2014 MU69 near the expected resolution 
limit.
That is, if during its formation process, accretion activity was sufficient to erase any craters
that may have been acquired during its assembly, MU69 will be modestly
cratered today.
It would thus be incorrect to conclude that a lightly cratered surface implies that
MU69's surface has been recently reset (or more extremely, that MU69 has been
recently created as a collisional fragment from a larger TNO).
Charon's extensive cryovolcanic plain (informally named Vulcan Planitia) is an example of this; it is unsaturated and the observed crater densities 
correspond to those expected from bombardment for $\approx4$ billion years
\citep{Singeretal2019}.

Because of this light cratering, MU69 exists in an environment very different
from main-belt asteroids of comparable ($\sim$10~km) size.
The latter are thought to essentially all be fragments produced after the
accretionary epoch, and thus are not `primordial' objects directly.
In contrast, MU69 will now look very much like it did at its formation
epoch (in the sense that impact processes have not made global structural
changes to the body nor even greatly affected more than a modest fraction
of the surface).
Despite MU69's low gravity, even the largest craters that form are
unlikely to provide enough distant ejecta to erase 200~m craters; 
ejecta dispersal will be even more limited if MU69 is very porous,
as seen at asteroid Mathilde \citep{Veverkaetal1999}.

Looking beyond the values of the crater densities themselves, no solar system body has ever
been  studied where the majority of the primary cratering is from
significantly subsonic projectiles.
This could result in craters with lowered depth/diameter ratios as seen for secondary craters \citep[e.g.,][]{BierhausSchenk2010},
and perhaps produce a morphologically visible difference between craters
formed by impactors coming in from the cold-classical population versus
those which tend to be in the 3-4 km/s range.
Seeing such differences will require well-resolved craters, which we expect 
to be modest in number even in the best MU69 images.
For the slowest impact speeds, we do not expect crater formation at all, but rather, accretion of the impactor material, either as coherent mounds or dispersed debris fields (``paintball" patterns).
In principle these speed differences might be able to help distinguish
between craters (or other features) formed by cold classical (kernel or stirred) impactors
versus those from the other sub-populations.

The uncertainty in the details of the dynamical state of the early outer
Solar System is only a small concern to our interpretation, due to the
dominance of the cold component's impactors.
Under a scenario that all the
hot populations were 100 times more populous for a brief ($\sim50$~Myr) 
of Solar System history (and a 100$\times$ enhancement
throughout that entire interval is unlikely), this would still only
double hot-population contributions to the cratering rate; Table~\ref{Table:impact_rate}
shows that even in this case the cold classical projectiles still dominate and
the total cratering only rises by a few tens of percent.
This is insufficient to qualitatively alter our conclusions that
MU69 will be dominated by subsonic projectiles and modestly
cratered.
If the object is heavily cratered, our interpretation would be that this is a
crater field preserved from the initial assembly epoch of MU69 
itself during the phase in which it accreted from smaller 
components.

Lastly, and most exciting, if the crater density and diameter distribution
are consistent with bombardment over the age of the Solar System by
the projectile distribution we derive based on the Pluto and Charon
cratering results, this will serve as convincing proof of a very shallow
sub-km size distribution in the Kuiper Belt itself (or, at least, in
the cold population which dominates the MU69 crater production 
rate).
Although 
\cite{Singeretal2019} 
already argue against the idea that surface geology on Pluto and Charon
could create the shallow size distribution near and below the elbow scale (a 
crater size distribution which is similar across a variety of terrains
on both bodies), 
such crater degradation ideas would be extremely unlikely to work in the
same way on tiny MU69.
A signature with crater $R$ values $\sim$0.001--0.01 that are increasing as 
one moves to larger crater diameters would confirm that the elbow
and its slope as seen on Pluto/Charon are due to a transition
to a shallow projectile size distribution.
If so, then not only is MU69 and its surface in a largely primordial
state, but the size distribution of the projectiles themselves is also
a largely un-evolved relic of the formation epoch.
This unusual state is possible because the $\alpha\simeq0.15$ distribution
is so shallow there are not enough projectiles to disrupt the rapidlly
shrinking targets as one moves down the size distribution.
The current Kuiper Belt size distribution (at least of the cold population)
would then be that which planetesimal accretion models would have to
create, and the cratering record preserved on the already-imaged objects
is directly providing constraints on the planet-building process.

%
%


\section{Acknowledgements}
\label{Section:acknowledgements}

S. Greenstreet acknowledges support from the B612 Foundation.
B. Gladman acknowledges support from the Canadian Natural Sciences and Engineering Research Council.
W.B. McKinnon, J.J. Kavelaars, and K.S. Singer acknowledge support from the New Horizons project.

\bibliographystyle{elsarticle-harv}
\bibliography{Bibliography.bib}

\end{document}